\documentstyle[preprint,aps]{revtex}
\tightenlines
\def\be{\begin{equation}}
\def\ee{\end{equation}}
\def\bea{\begin{eqnarray}}
\def\eea{\end{eqnarray}}
\def\ba{\begin{array}}
\def\ea{\end{array}}

\def\d{\delta}
\def\e{y}
\def\x{x}

\def\0{$\Gamma_0$}
\def\l{u^*}
\def\o{\omega}

\def\L{${\cal L}$ }

\begin{document}
\draft

\title{On the duality relation for  correlation functions 
of the  Potts model}
 
\author{Wentao T. Lu and F. Y. Wu \\
Department of Physics\\
 Northeastern University, Boston,
Massachusetts 02115}

\maketitle

\begin{abstract}
We prove a recent conjecture on the duality relation for correlation
functions of the Potts model
 for  boundary spins of a planar lattice.
Specifically, we deduce
the explicit  expression for
the duality of the $n$-site correlation functions,
and 
establish  sum rule identities 
in the form of the M\"obius inversion
of a partially ordered set. 
The strategy of the proof is by first formulating the problem
for the more general
 chiral Potts model. The extension of our consideration
to 
 the many-component  
Potts models is also given.

 \end{abstract}

\vskip 1cm
\pacs{05.50.+q}

\section{Introduction}
The duality relation for the $q$-state Potts model 
 \cite{potts,wuPotts}
is an identity
 \cite{wuwang}  relating the partition functions of a Potts model
on a planar lattice with that of its dual.
Very recently,
this  duality
consideration has been extended  to the 
Potts  correlation functions \cite{wu,wuhuang}.  Specifically, 
it was established that
 certain  duality relations exist 
for correlation functions of $n$ 
Potts spins   on the boundary of a planar lattice.
Explicit expressions for the duality relation have been obtained for
$n=2,3$ \cite{wu} and $n=4$ \cite{wuhuang}; the expression for  
 general $n$ has
also been conjectured \cite{wuhuang}.
It has also been shown that there exist
certain sum rule identities, but the explicit form of the
identities was not given \cite{wuhuang}.
The purpose of this paper is the following:  We first
extend  the consideration
to  the chiral Potts model \cite{chiral} and obtain its
duality relation in a very  general form.
 We then show that this  formulation
 permits us to establish the conjecture of \cite{wuhuang}.  
In addition, it also leads to an expression of the sum rule identities
in the form of  the M\"obius inversion of a partially ordered set.
Furthermore,  starting from 
a  multi-component chiral Potts model \cite{perk}, we
 extend considerations to  the multi-component  Potts model, 
which includes
 the Ashkin-Teller model \cite{at} and
its  generalization \cite{domany}.
This extends  
results  reported elsewhere  \cite{wulu} for the 2-component Potts model.

\section{The chiral Potts model}
Consider a $q$-state spin system on a two-dimensional lattice
or, more generally, any planar graph ${\cal L}$.
 Let $\x_i=1,2,\cdots, q$ denote the spin state of the $i$-th
site and $-J(\x_i, \x_j)$ the interaction energy between sites $i$ and $j$,
which can be edge-dependent.
The  interaction is chiral if
\be
J(\x, \x') \not= J(\x', \x). \label{chiralinter}
\ee
The partition function of this spin system is
\be
Z(u) =\sum^{q}_{\x_i=1} \prod_{<ij>} U(\x_i, \x_j), 
\ee
where $U(\x, \x')=\exp[J(\x, \x')/kT]$ and the product is taken over
all edges of ${\cal L}$.

For our purposes we consider  the chiral Potts model
\cite{chiral} for which the
Boltzmann factor is cyclic, namely, it satisfies
 \be
U(\x, \x')=u(\x-\x'),\quad \quad ({\rm mod} \quad q).\label{chiralpotts}
\ee
We shall refer to the $q\times q$ matrix ${\bf U}$ 
as the interaction matrix.
    The chiral Potts model reduces to the standard Potts model
upon taking   $J(\x, \x')/kT =K \d(\x, \x')$, where $\d$ is the 
Kronecker delta function which we define   as
\bea
\d(\x, \x') &=& 1, \hskip 1.5cm \x=\x' \hskip0.5cm ({\rm mod}\>\>q) \nonumber\\
  &=& 0, \hskip 1.5cm \x\not=\x' \hskip0.5cm ({\rm mod}\>\>q). 
\eea
This leads to  
\begin{equation}
u(\x) = 1 +(e^K-1)\d(\x, 0) .
 \label{standardpotts}
\end{equation}

{\it Duality relation for the partition function}:
The partition function of
the chiral Potts model
 possesses a duality relation  \cite{wuwang}.
  Let  ${\cal L}^D$ be  the dual of ${\cal L}$,
  and let $Z^{(D)}(\l)$ be the partition function of the chiral Potts model
on ${\cal L}^D$
with  the Boltzmann factor\footnote{The definition of the dual Boltzmann factor 
adopted here differs
from that used in \cite{wuwang} by a factor $\sqrt q$.  
The present definition is consistent to the notion of $(u^*)^* =u$.}
\be
\l(\e) =  {1\over {\sqrt q}}
\sum_{\x=1}^{q}\> \o^{\x\e}\> u(\x), \quad\quad \e=1, 2, \cdots, q \label{eigen}
\ee
where $
\o= e^{2\pi i/q}$. 
 Then, the duality relation      reads \cite{wuwang}
\be
q^{-N/2}Z[u( \x)] = q^{-N^*/2} Z^{(D)}[u^*( \e)],  \label{duality}
\ee
where  $N$ and $N^*$ are, respectively, the numbers of
sites of \L and  ${\cal L}^D$.  They are related 
to the number of edges E by the Euler relation
$N + N^* = E +2. $
 This duality relation holds for arbitrary edge-dependent interactions.
Note that $\sqrt q u^*(\e)$, $\e= 1,\cdots,q$, are the eigenvalues of
the matrix ${\bf U}$.  In the case of the standard Potts model,
(\ref{eigen})  reduces to
\begin{equation}
u^*(\e) = {1\over \sqrt q} \biggl[e^K -1 +q\>\d(\e,0)\biggr].
\end{equation}

\section{Duality relation for the correlation function}
Number $n$ sites on the boundary of \L by integers 
 $1,2,\cdots n$   clockwise  as shown in Fig. 1.
The probability that the $n$ sites are in respective spin states
$\x_1, \x_2, \cdots, \x_n$ is 
\be
P_n(\x_1, \x_2, \cdots, \x_n) = Z_{\x_1\x_2\cdots\x_n} / Z(u),  \label{prob}
\ee
where $Z_{\x_1\x_2\cdots\x_n}$ is the {\it partial} partition function
of the chiral Potts model  with the $n$ 
 spins  in fixed given states.
 Construct an auxiliary lattice ${\cal L}_{\rm aux}$  
 from ${\cal L}$  by 
   connecting all $n$ boundary sites to an extra  site $e$
in the infinite face of ${\cal L}$ as shown.
These $n$ connecting lines then divide the infinite face of \L into
$n$ parts, which we also    number by $1,2,\cdots, n$ as shown in Fig. 1.
  Let $u_{je}$, $j=1,2,\cdots n$, be the Boltzmann factor of the line
 connecting sites $j$ to $e$.  Its dual edge 
then connects  sites (residing in faces of ${\cal L}$) $j$ and $j+1$ on
 ${\cal L}^D_{\rm aux}$
  with the Boltzmann factor 
 \be
\l_{j+1,j}(\e_{j+1}-\e_j) = 
{1\over \sqrt q}\sum_{\x=1}^{q} \>
\o^{(\e_{j+1}-\e_j)\x}
\> u_{je}(\x), \hskip 1cm j=1,\cdots ,n
 \label{eigen1}
\ee
where  $y_j$ denotes the spin state of the dual site $j$.  Here we have adopted the 
convention of \cite{wuwang} in orienting the  edges.
 It is also convenient to consider the $n$-spin correlation \cite{wu}
\be
\Gamma_n = q^n P_n(x,x,\cdots,x) -1, \label{g1}
\ee
a quantity which vanishes identically if the $n$ spins are completely uncorrelated.

We  apply the duality relation (\ref{duality}) to ${\cal L}_{\rm aux}$.
  Let 
$Z^*_ { \e_1 \e_2\cdots\e_n}$ be the partial partition function
of the Potts model  on ${\cal L}^D_{\rm aux}$ with  sites $1,2,\cdots , n$
in respective states
$\e_1, \e_2,\cdots,\e_n$.
Then, explicitly,
(\ref{duality}) reads 
\bea
&& q^{-(N+1)/2} \sum_{\x_1\x_2\cdots\x_n} \sum_{\x_e} 
 u_{1e}(\x_{1e}) 
u_{2e}(\x_{2e})\cdots u_{ne}(\x_{n e}) Z_{\x_1 \x_2\cdots\x_n} \nonumber \\
&& \hskip 2cm =q^{-(N^*+n-1)/2}   
\sum_{\e_1\e_2\cdots \e_n} 
 \l_{21} ( \e_{21})
\l_{32}(\e_{32})\cdots \l_{1n}(\e_{1n})
Z^*_ { \e_1 \e_2\cdots\e_n}, \label{ndual}
 \eea
where $\x_{1e} = \x_1 - \x_e$, $\e_{21} = \e_2 - \e_1$, etc.,
and we have used the fact that
 ${\cal L}_{\rm aux}$ has $N+1$ sites and
${\cal L}^D_{\rm aux}$ has $N^*+n-1$ sites.\footnote{Note that the partial partition function
  $Z^*_{y_1y_2\cdots y_n}$ in (\ref{ndual}) 
differs from that in \cite{wu,wuhuang} by
a  factor.}

Due to the cyclic nature of the matrix elements, we can 
 replace the arguments
$\x_{1e}, \x_{2e},
\cdots$  on the LHS of (\ref{ndual})
by $\x_1, \x_2, \cdots$.  The summation
over $\x_e$ can then be carried out 
leading to an overall factor $q$.   Next we
eliminate $u^*$  by
 using (\ref{eigen1}).   Since all $u$ factors in  (\ref{ndual})
are independent,  as a consequence
we can equate the coefficients of 
$ u_{1e}(\x_{1}) 
u_{2e}(\x_{2})\cdots u_{ne}(\x_{n })$ 
in  (\ref{ndual}).   This leads to the duality identity
 \be
Z_{\x_1 \x_2\cdots\x_n}= C_n(q)\sum_{\e_1 \e_2\cdots \e_n} 
{ M}(\x_1, \x_2,\cdots,\x_n \>
|\> \e_1, \e_2,\cdots,\e_n)
Z^*_ { \e_1 \e_2\cdots\e_n}, \label{gen01}
\ee
where 
\be
C_n(q) = q^{-n+(N-N^*)/2},
\ee
 and 
\be
 {M}(\x_1, \x_2,\cdots,\x_n \>  |\> \e_1, \e_2,\cdots,\e_n)
      =\o^{\x_{n1}\e_1+
\x_{12}\e_2+\cdots+\x_{(n-1)n}\e_n} \label{transfer}
 \ee
can be interpreted as the elements of a $q^n\times q^n$ matrix {\bf M}.
Eq. (\ref{gen01}) is the  duality relation for the chiral Potts model,
and is a very general expression.  The relation assumes a
 reduced  form for the    standard Potts model
after taking into account the degeneracy of states, and the crux of matter is,
of course, to deduce  the reduced 
relation. This is the subject matter of the next section.

Introducing the notation
\be
p_{y_1y_2\cdots y_n} \equiv 
   Z^*_{y_1y_2\cdots y_n}/Z^*_{11\cdots 1} =q Z^*_{y_1y_2\cdots y_n}/Z^{(D)}\label{p}
\ee
and combining with (\ref{duality}) and (\ref{prob}),  we can
rewrite  (\ref{gen01}) as
 \be
P_n(\x_1, \x_2,\cdots,\x_n)= q^{-(n+1)}\sum_{\e_1 \e_2\cdots \e_n} 
{ M}(\x_1, \x_2,\cdots,\x_{n-1},\x_n \>
|\> \e_1, \e_2,\cdots,\e_n)
p_ { \e_1 \e_2\cdots\e_n}. \label{gen03}
\ee
An  immediate consequence of (\ref{gen03}) is
\be
\Gamma_n = {1\over q} \sum_{\e_1 \e_2\cdots \e_n} 
  p_ { \e_1 \e_2\cdots\e_n}-1,  \label{gamman}
\ee
a result which has been reported
 previously \cite{wuhuang,jacobsen}.

To obtain the inverse of (\ref{gen01}), we note that,
 by deleting the $\{1,2\}, \{2,3\}, \cdots $ edges 
 in ${\cal L}^D_{\rm aux}$, we arrive 
at   a lattice ${\cal L}^*$
in which  $1,2, \cdots$ of ${\cal L}^*$ are  its boundary sites.
An example of \L and the resulting  ${\cal L}^*$ is  shown in Fig. 2. 
 It is clear  that  the roles of \L and ${\cal L}^*$ are 
reciprocal,  so we can apply (\ref{gen01}) to
  ${\cal L}^*$ to obtain an inverse transform.
Now  ${\cal L}^*$ has ${\bar N}=N^*+n-1$ sites and 
its dual  has ${\bar N}^*=N-n+1$ sites.  
 As a result,  we obtain 
 \be
Z^*_ { \e_1 \e_2\cdots \e_n}=  {\bar c}_n(q) \sum_{\x_1 \x_2\cdots \x_n} 
{ M}(\e_1, \e_2,\cdots,\e_n\>|\> \x_n, \x_1,
\x_2,\cdots,\x_{n-1})
Z_{\x_1 \x_2\cdots\x_n},  \label{inverse}
\ee
where 
\be     
{\bar c}_n(q)   = q^{-n +({\bar N} - {\bar N}^*)/2} = q^{-1+(N^*-N)/2}.
\ee
We have $c_n(q){\bar c}_n(q) = q^{-(n+1)}$
and note that the sequence  $\{\x_n, \x_1, \x_2, \cdots, \x_{n-1}\}$
of the indices on the RHS
is related to  $\{\x_1, \x_2, \cdots, \x_{n-1}, \x_n\}$
by a simple cyclic permutation. 
 Upon
combining (\ref{inverse}) with (\ref{gen01}) we obtain the
identity
\be
{\bf M}^2(\x_1, \x_2,\cdots,\x_n \>
|\> \x'_{1}, \x'_{2},\cdots,\x'_{n})
=q^{n+1} \d({\x_1,\x'_{2}})\d({\x_2,\x'_{3}})
\cdots\d({\x_n,\x'_{1}}), \label{mm2}
\ee
which we refer to as a  reciprocal inversion relation.

\section{The standard Potts model}
For the standard Potts model  the
duality relation (\ref{gen01}) assumes a reduced  form which
 has been given previously   
 for $n=2,3$ \cite{wu}
and for $n=4$ \cite{wuhuang}.
  The form for general $n$
has  also been conjectured \cite{wuhuang}. It has also been shown that
the correlation functions satisfy certain sum-rule identities \cite{wuhuang}.
  Here, we reformulate 
the standard Potts 
model as an instance of the chiral Potts model and,
as we shall see, this   leads us to 
  establish the conjecture as well as deduce the general expression for the 
sum rule identities.
  For this reason,
it is instructive to first demonstrate how   (\ref{gen01}) reduces to the
known results  for $n=2,3$.   

The standard Potts model is characterized by the fact that one needs
only to keep track of  spin states that are  the same.
That is to say, one needs to keep track of the partition of the
$n$ sites into {\it blocks}, such that sites in one block are
in the same state.  This leads to writing  the partial partition
functions in an  expansion of a partially ordered set.
Along this line, we writes for $n=2,3$,
 \bea
Z_{\x_1\x_2} &=& D_{12} + D_{11} \d(\x_1, \x_2) \nonumber  \\
Z^*_{\e_1\e_2} &=& D^*_{12} + D^*_{11} \d(\e_1, \e_2) \nonumber\\
Z_{\x_1\x_2\x_3}&=& D_{123} +D_{113} \d(\x_1,\x_2) +D_{121} \d(\x_1,\x_3) +
 D_{122} \d(\x_2,\x_3) + D_{111} \d(\x_1,\x_2,\x_3) \nonumber \\
Z^*_{\e_1\e_2\e_3}&=& D^*_{123} +D^*_{113} \d(\e_1,\e_2) +D^*_{121} \d(\e_1,\e_3) +
 D^*_{122} \d(\e_2,\e_3) + D^*_{111} \d(\e_1,\e_2,\e_3), \label{n32}
\eea
 where $\d(\x_1,\x_2,\x_3) = \d (\x_1, \x_2)\d(\x_2, \x_3)$.  
Here, for $n=3$, for example,
the partition  of the 3-set of integers $\{1,2,3\}$ includes the 5 elements
$\bigl\{ \{1\},\{2\}, \{3\} \bigr\}$, $\bigl\{ \{12\}, \{3\} \bigr\}$,
$\bigl\{ \{1\},\{23\} \bigr\}$, $\bigl\{ \{13\},\{2\} \bigr\}$,
$\bigl\{ \{123\}\bigr\}$  denoted by  subscripts $\{123\}$, $\{113\}$, $\{122\}$,
$\{121\}$, $\{111\}$, respectively.

For $n=2$, we substitute the first two lines of  (\ref{n32}) 
into (\ref{gen01}).   Making use of the identity
  \be
\sum_{\e=1}^q \o^{xy} = q \> \d(\x,0), \label{sum}
\ee
  one arrives at  
\bea
 D_{12} + D_{11} \d(\x_1, \x_2) 
&=& C_2(q) \sum_{\e_1\e_2}\o^{\x_{21}\e_1
+ \x_{12}\e_2}\biggl[D^*_{12} + D^*_{11}\d(\e_1, \e_2) \biggr] \nonumber \\
&=&
 C_2(q) \biggl[ q^2 D^*_{12} \d(\x_1,\x_2) + q D^*_{11}\biggr]. \label{n222}
\eea
 Since this equation holds for arbitrary $x_1$ and $x_2$, 
 the coefficients of corresponding delta functions must be equal, and this leads 
to the identities
 \be
D_{12} = C_2(q) \>q D_{11}^*, \hskip 1cm D_{11} = C_2(q) \>q^2 D_{12}^*,\label{d2}
\ee
which  can further be converted to   relate $Z_{x_1x_2}$ to
$Z^*_{x_1,x_2}$ as follows.  First, the second line of (\ref{n32}) gives
\be
Z^*_{12} = D^*_{12}, \hskip 2cm Z^*_{11} = D^*_{12} + D^*_{11},\label{d21}
\ee
from which one  obtains the inverse
 \be
D^*_{12}=Z^*_{12}, \hskip 2cm D^*_{11} = Z^*_{11} - Z^*_{12}. \label{d22}
\ee
Substituting (\ref{d22}) into (\ref{d2}) and combining
 (\ref{prob}), (\ref{duality}) and the definition (\ref{p}),
  one is led to the duality relation for the 2-point correlation function
\bea
P_{2}(x_1, x_2) &=& [D_{12} +D_{11} \d(x_1, x_2)] /Z \nonumber \\
 &=&{1\over {q^2}}\biggl[ 1+\biggl(q\d(x_1,x_2) -1\biggr) p_{12} \biggr].
\label{p2} 
\eea
This is the duality relation (\ref{gen03}) for $n=2$.
Finally, using (\ref{n32}), (\ref{d2}), and (\ref{d22}), and introducing the
row vector ${\tilde {\bf z}}_2 =(Z_{11}, Z_{12})$, ${\tilde {\bf z}}^*_2 
=(Z^*_{11}, Z^*_{12})$, we find 
\be
{\bf z}_2 = qC_2(q){\bf T}_2(q) \cdot {\bf z}^*_2, \label{t2}
\ee
where
\be
 {\bf T}_2(q) = \pmatrix{1 & q-1 \cr 1 & -1\cr},\label{tt2}
\ee
satisfying $[{\bf T}_2(q)]^2 = q{\bf I}_2$,
with ${\bf I}_2$ being the $2 \times 2$ 
identity matrix.

Similarly for $n=3$ one obtains in place of (\ref{n222})
\bea
 D_{123} &+&D_{113} \d(\x_1,\x_2)+D_{121} \d(\x_1,\x_3) +
 D_{122} \d(\x_2,\x_3) + D_{111} \d(\x_1,\x_2,\x_3) \nonumber \\
&=& \hskip 0.5cm C_3(q) \sum_{y_1 y_2 y_3} \>
 \o^{ x_{31}y_1+x_{12}y_2+x_{23}y_3}
\biggl[ D^*_{123} +D^*_{113} \d(\e_1,\e_2) +D^*_{121} \d(\e_1,\e_3)\nonumber \\
 &&\hskip 2cm +
 D^*_{122} \d(\e_2,\e_3) + D^*_{111} \d(\e_1,\e_2,\e_3)\biggr], \label{dd3}
\eea
 leading to the identities
\bea
D_{123} &=& C_3(q) \>q D^*_{111}, \hskip 2cm
D_{111} = C_3(q)  \>q^3 D^*_{123}  \nonumber \\
D_{113}& =&C_3 \>q^2 D^*_{121}, \quad 
D_{121} = C_3(q) \>q^2 D^*_{122}, \quad 
D_{122} = C_3(q) \>q^2 D^*_{113}. \label{d32}
\eea
Again, we can rewrite (\ref{d32}) in a form  relating
$Z_{x_1x_2x_3}$ to $Z^*_{y_1y_2y_3}$.
Now the inverse of the last line of (\ref{n32}) is
\bea
D^*_{123} &=& Z^*_{123} \nonumber \\
D^*_{113} &=& Z^*_{113} - Z^*_{123}, \hskip .5cm
D^*_{121} = Z^*_{121} - Z^*_{123}, \hskip .5cm
D^*_{122} = Z^*_{122} - Z^*_{123}, \nonumber \\
D^*_{111} &=& Z^*_{111} -(Z^*_{113} +Z^*_{121} +Z^*_{122}) + 2 Z^*_{123}.\label{d33}
\eea
Substituting (\ref{d33}) into (\ref{d32}) and 
using the third line of (\ref{n32}) for $Z_{x_1x_2x_3}$,
we obtain 
\bea
P_3(\x_1, x_2, x_3) &=& {1\over {q^3}} \biggl[ 1-(p_{122}+p_{121}+p_{113})+2 p_{123}
   +q(p_{121} -p_{123}) \d(x_1, x_2)\nonumber \\
&& +q(p_{122} -p_{123}) \d(x_1, x_3) 
 +q(p_{113} -p_{123}) \d(x_2, x_3) +q^2 p_{123}\d(x_1,x_2,x_3) \biggr]. \label{p3}
\eea
This is the duality relation (\ref{gen03}) for $n=3$ reported in
\cite{wu}.\footnote{Note, however, the numbering of $\{y_1,y_2,y_3\}$
as $\{s', s'', s\}$ in \cite{wu}, resulting in a more symmetric   
appearance of (\ref{p3}).}

 Explicitly, introducing the
row vector ${\tilde {\bf z}}_3 =(Z_{111}, Z_{113},Z_{121}, Z_{122},Z_{123})$
and similarly defined ${\tilde {\bf z}}^*_3$,
  we find 
\be
{\bf z}_3 = qC_3(q){\bf T}_3(q) \cdot {\bf z}^*_3,  \label{t3}
\ee
where
\be
 {\bf T}_3(q) = \pmatrix{1 & q-1& q-1& q-1 &(q-1)(q-2)\cr 1 & -1& q-1 & -1 & -(q-2)\cr
 1 & -1 & -1 & q-1 & -(q-2)\cr 1 &q -1  &-1 & -1 & -(q-2)\cr
1&-1&-1&-1&2} \label{tt3}
\ee
satisfying
\be
[{\bf T}_3(q)]^2 = q^2\pmatrix{ 1 & 0 & 0 &0 &0 \cr 0 & 0 & 0 &1 &0 \cr 0 & 1 & 0 &0 &0 \cr
0 & 0 & 1 &0 &0 \cr 0& 0 & 0 &0 &1 \cr}.
\ee
 
\section{The general analysis} 
For general $n$, we write
 $Z_{\x_1\cdots\x_n}$ 
 as a  sum in the form of  (\ref{n32}), namely,
\be
Z_{x_1\cdots x_n} = \sum_{X} D_{X} \d(X),
  \label{Zn1}
\ee
where the summation is over all partitions $X$ of $n$ integers
$\{1,2,\cdots, n\}$,
and $\d(X)$ is a  product of        
 delta  functions  associated with  blocks in  $X$.
 For example, we have $\d(X)=\d(x_1,x_2)\d(x_3,x_4)$ for $X=\{1133\}$,
and  $\d(X)=\d(x_1,x_2,x_3)=\d(x_1,x_2)\d(x_2,x_3)$ for $X=\{1114\}$.
If a block of $X$ contains 
$m\geq 3$ $x$'s as in $\d(x_1,x_2,x_3)$ above, we 
write $\d(X)$ as a product of $m-1$  2-point delta functions $\d(x_j, x_k)$ 
with the $x$'s arranged in, say, a clockwise
sequence, so the writing of $\d(X)$ as a product of 2-point delta functions
is unique.   
The number of terms in (\ref{Zn1})  is $b_n$, and
it has been shown in \cite{wuhuang} that $b_n$ is generated by
\be 
\sum_{n=0}^\infty b_n t^n /n! = {\rm exp} (e^t-1) .
\ee
Without the loss of generality we shall assume $q\geq n$
so that all  $Z_X$ are physically
realized.\footnote{Otherwise
the situation  $q<|X|\leq n$ can occur, where $|X|$ is      the number
of blocks of the partition $X$, and $Z_X$ has no physical meaning. 
 But  $Z_X$, which can be computed
for  $q\geq |X|$, is a polynomial in $q$.
We shall take   $Z_X$ to be the same polynomial for all $q$
including $q<|X|$.
With this understanding  in mind all our results, including the identities derived
below, hold for all $q,n$.}
   
The partition sums (\ref{n32}) and (\ref{Zn1}) can be regarded as a transformation 
between the $Z$'s and $D$'s. 
The transformation 
 is that of a 
partially ordered set and it is known \cite{mobius} that
the inverse  is
   given by the M\"obius inversion.
Let $X, X', Y,$ etc.  denote  specific 
partitions of an $n$-set.
 We write $  X' \preceq X$ if the block structure of $X'$ is contained in
 $X$, namely, $X'$ is a  refinement of $X$ \cite{mobius}.
  For example, $\{1134\}$ and $\{1234\}$ are contained in
$\{1133\}$, while $\{1224\}$ is not.
Then, an immediate consequence of (\ref{Zn1}) is \cite{mobius}
 \be
Z_X= \sum_{X'\preceq X} D_{X'}. \label{Zn2}
\ee
In addition,  the M\"obius inversion of (\ref{Zn1}) is
\be
D_{X} =\sum_{X' \preceq\> X} \mu(X', X) Z_{X'} \label{mobius}
\ee
with
\be
\mu(X', X) = (-1)^{|X'| -|X|} \prod_{{\rm blocks\>} \in X}(n_b-1)!,  
\ee
 where $|X|$ and $|X'|$ are, respectively, the number of blocks in $X$
and $X'$, and 
$n_b$ the number of blocks of $X'$ that are contained in a block of $X$.
 In the examples  above, for instance, 
 $X=\{1133\}$ has two blocks, and we have
$n_b= \{1,2\}$ for  $X'=\{1134\}$, and
$n_b= \{2,2\}$ for  $X'=\{1234\}$, etc.   This leads to
\bea
D _{1133} &=& (-1)^{2-2}(0!)^2 Z_{1133}+(-1)^{3-2} (0!)(1!) Z _{1134}
+(-1)^{3-2} (1!)(0!) Z _{1233}  + (-1)^{4-2} (1!)^2 Z _{1234}  \nonumber \\
&=& Z _{1133} - Z _{1134} - Z _{1233} + Z _{1234}. \nonumber \\
\eea
Other examples are
\bea
 D _{1114} &=& Z _{1114} - Z _{1134} -Z _{1214} - Z _{1224} + 2 Z _{1234}\nonumber \\
D _{1212} &=& Z _{1212} -Z _{1214}-Z _{1232} +Z _{1234}. \label{mobius1}
\eea 
Substitute  the partition sum (\ref{Zn1}) and 
a similar expression for
$Z^*_{y_1\cdots y_n}$  
 into (\ref{gen01}), we obtain
\be
\sum_{X} D_{X} \d(X) =C_n(q)\sum_{Y} 
\sum_{y_1\cdots y_n} M(x_1,\cdots x_n|y_1,\cdots y_n) D^*_{Y } \d(Y).  \label{sum1}
\ee 
 
It turns out that the analysis is best done graphically.
Represent  the matrix element $M(x_1,\cdots,x_n | y_1, \cdots, y_n)$
 in (\ref{gen01})
by a graph $G_n$ shown in Fig. 3, where each   node (open circle) denotes
a  $\e$-summation,  and each 
arc  is assigned a label $\x$ such that
an outgoing arrow (from a node labelled $y$)   carries a factor $\o^{-\x\e}$ 
and an incoming arrow  a factor $\o^{\x\e}$.
To each  $X$ we construct a ``connectivity"
$\Gamma_X$ according to 
 the following prescription:
  Connect the
mid-points of the arcs belonging to  each block of $X$ to a
common point exterior to $G_n$. 
   A connectivity is planar if the connecting lines
do not intersect; otherwise the connectivity is non-planar.
Examples of connectivities for $G_4$ are shown in Fig. 4.
 We note that the construction of the connectivities   is the same as that introduced
in \cite{wuhuang}.
 We have two possibilities.

i) $\Gamma_X$ is planar.
  In this case  $\Gamma_X$ 
divides the region exterior to $G_n$ into {\it faces}.  Regarding all nodes
in one face as belonging to a block of another partition $Y$, 
 and denote this mapping by
  $X\to Y$.  It is clear that
  $\Gamma_Y$
is also planar and that  $Y\to X$.  Therefore   
  the mapping  $X\leftrightarrow Y$  is  one-one.
  One has also\footnote{For $X=\bigl\{\{x_1\},\{x_2\},\cdots, \{x_n\} \bigr\}$, $|X|=n$,
we have $Y=\big\{ \{y_1, y_2,\cdots, y_n \} \bigr\}$, $|Y|=1$.
   It is also clear that $|Y|$ increases by 1 when $|X|$
decreases by 1.  This establishes (\ref{xy}).} 
\be 
|X| + |Y| = n+1. \label{xy}
\ee
It has been shown \cite{temperleylieb,bn} that the number $c_n$ of planar $\Gamma_X$,
or simply planar $X$, is generated by
\be
\sum_{n=0}^\infty c_nt^n = (1-\sqrt {1-4t})/2t.
\ee

With these understandings, we now carry out the $y$-summations on the RHS
of (\ref{sum1}). From (\ref{sum}) 
each $y$-summation yields a factor $q\>\d(\x,\x')$,
where $x$ and $x'$ are the labellings of the incoming and outgoing arrows.
 One can then regard the labellings $x$  as  a flow
 which is conserved at each node.
Further, the effect of the
delta function $\d(Y)$ in (\ref{sum1}) is to  collapse the $y$-summations
of each block of $Y$ into a single summation,
  resulting in a factor $q\>\d(x_{\rm in}, x_{\rm out})$
where $x_{\rm in}$ is the sum of the incoming flows and $x_{\rm out}$
the sum of outgoing flows. 
 This effect can be  conveniently visualized by contracting the nodes in 
question into a single one, and requiring that
the flow is conserved at the contracted node.
 For example, as shown in Fig. 5, the partition $Y=\{123216\}$ of  $G_6$
is contracted  into a diagram consisting
of a sequence of  3 $G_2$ ``cactus leaves" attached together  at 
two common nodes.
  Generally, the delta function $\d(Y)$ 
contracts $G_n$ into a  diagram consisting of cactus leaves
 $G_m, m<n$, attached together  at common nodes.
The planar nature of $Y$ then ensures that the resulting diagram is a cactus {\it tree}.
   One next  carries out
the $y$-summations one-by-one starting from the nodes of the outermost leaves
of the tree.
The tree structure now  ensures that this
  always leads to a product of 2-point delta functions. 
In the example shown in 
  Fig. 5, for instance, this leads to
\be
\d(x_5,x_6)\d(x_4+x_6,x_1+x_5)\d(x_1+x_3,x_2+x_4)\d(x_2,x_3)
=\d(x_5,x_6)\d(x_4,x_1)\d(x_2,x_3).  \label{delta}
\ee
But this is precisely the factor $\d(X)$ appearing in 
 $D_X\d(X)$ on the LHS of (\ref{sum1}), where $X
= \{122155\} \leftrightarrow  Y$.  
It is readily verified that, generally for each planar $X\leftrightarrow Y$, to the term $D_X\d(X)$ on the LHS of (\ref{sum1}), 
 the $y$-summations of the $Y$ term on the RHS
yields a factor $D_Yq^{|Y|}\d(X)$.
Now, the indices $\{x_i\}$ in a given $\d(X)$ are arbitrary.
It follows that  
the coefficients of the two terms must be equal, and this leads to the identity
\be
D_X = C_n(q) \>q^{|Y|}D^*_Y, \hskip 1cm {\rm planar\>\>}
 X\leftrightarrow Y,
    \label{duality1}
\ee
an equation which generalizes  (\ref{d2}) and (\ref{d32}).

ii) $\Gamma_X$ is non-planar.
We now consider the remaining $b_n-c_n$ terms in (\ref{sum1}).
Now, each remaining terms on the LHS is  of the form $D_X\d(X)$,
where $\d(X)$ is   a product of 2-point delta functions $\d(x_j, x_k)$.
For the remaining terms on the RHS we  again carry out the $y$-summations for
each $Y$.  However, due to the fact that $Y$ is now non-planar, the
process of contraction as described in the above
leads to  diagrams in which cactus leaves form rings, or  circuits. 
It follows that the $y$-summations  generates a product involving some
 4-point delta functions of
the type $\d(x_i+x_j, x_k + x_\ell)$,  which cannot be  reduced 
into  products of   2-point delta functions as in (\ref{delta}).
   In other words, 
the delta functions $\d(X)$ occurring on the LHS do not appear on
the RHS, and vice versa.  Now, the 
the indices $\{x_i\}$ in a given $\d(X)$ are arbitrary.
It follows that,  
for the equality (\ref{sum1}) to hold for all $\{x_i\}$, each of
these remaining coefficients must vanish individually.  Namely, we must have
\be
D_X = D^*_Y = 0, \hskip 1cm {\rm non-planar \>\>} X, Y.  \label{duality2}
\ee
   
Finally  introduce the partially ordered partition sums
\bea
P_n(x_1, x_2, \cdots, \x_n) &=& \sum_X A_X\d(X) \nonumber \\
p_{y_1y_2\cdots y_n}
 &=& \sum_Y B_Y\d(Y), \label{relation}
\eea
where $A_X=D_X/Z$, $B_X=Z^*_Y/Z^*_{11\cdots1}$. Combining (\ref{duality}),
(\ref{p}) and using
(\ref{xy}), the identities (\ref{duality1}) and (\ref{duality2}) now
establish the conjecture of \cite{wuhuang}, which we now state as a

\noindent
{\it Theorem}:
 \bea
A_{X} &=&  q^{-|X|} B_{Y},\hskip 1.8cm {\rm for \>\>planar\>\>}X \leftrightarrow Y,
   \nonumber  \\
  &=& 0, \hskip 3.1cm {\rm otherwise}.
     \label{conjecture}
\eea
Explicitly, the coefficients $B_Y$ are given in terms of     
 $p_{y_1y_2\cdots y_n}$ by  the M\"obius inversion (\ref{mobius}).
Thus,  the second line in (\ref{conjecture})
leads to  identities relating  partial partition functions $Z_X$.
  For example, from the second line of
(\ref{mobius1}) and $A_{1212}=D_{1212}=0$, we obtain
\be
Z _{1212} -Z _{1214}-Z _{1232} +Z _{1234}=0. \label{mobius2}
\ee
This is the $n=4$ identity  reported in \cite{wuhuang}.
Generally, one obtains an identity $A_X=0$ for each non-planar $X$.
Using these equations  one can  express all  non-planar $Z_X$ 
  in terms of planar ones, a fact reached in \cite{wuhuang} through the use of
 high-temperature expansions.
  
One can also write down the transformation relating $Z_X$ and $Z^*_Y$.
For general $n$ and planar $X$
one finds by combining (\ref{Zn2}), (\ref{duality1}) and (\ref{mobius}),
\be
Z_X  = C_n(q) \sum _{X' \preceq X} q^{|Y'|} \sum _{Y'' \preceq Y'}
\mu (Y'',Y') Z^*_{Y''}, \hskip 1cm X' \leftrightarrow Y'.\label{xxx}
\ee
The transformation relating $P_n(x_1,\cdots,x_n)$ to $p_{y_1\cdots y_n}$
can be written down similarly with the factor $C_n(q)$ replaced by $q^{-(n+1)}$.
More explicitly,
define a row vectors ${\tilde {\bf z}}_n$
and ${\tilde {\bf z}}^*_n$ whose
elements are the $c_n$  partial partition functions corresponding to planar $X$,
we find
\be
{ {\bf z}}_n = qC_n(q){\bf T}_n(q)  \cdot { {\bf z}}^*_n, \label{tn}
\ee
where ${\bf T}_n(q) $ is a $c_n \times c_n$ matrix satisfying the identity
\be
[{\bf T}_n(q)]^2 (X,X') = q^{n-1}
 \d({\x_1,\x'_{2}})\d({\x_2,\x'_{3}})
\cdots\d({\x_n,\x'_{1}}).
\ee
Note that the coefficient $q^{n-1}$ on the RHS differs from that
in (\ref{mm2}).
 Expressions of ${\bf T}_n(q)$ 
for $n=2,3$ have been given in (\ref{tt2}) and (\ref{tt3}).
The expression for $n=4$ can be deduced from results reported in  \cite{wuhuang}.
 For $n=5$ there are 42 planar $X$ and we    define the
 row vector
\bea
&&{\tilde {\bf z}}_5 =\biggl( Z_{11111}, (Z_{11115},Z_{11141},Z_{11311},Z_{12111},Z_{12222}),(Z_{11144},Z_{11331}, Z_{12211}, Z_{11333},Z_{12221}),\nonumber \\
 &&\hskip .5cm       (Z_{11145},Z_{11341}, Z_{12311}, Z_{12333}, Z_{12225}),
        (Z_{11335},Z_{12241}, Z_{11344}, Z_{12331}, Z_{12244}),\nonumber \\
     &&\hskip .5cm   (Z_{11315},Z_{12141},Z_{12311}, Z_{12115}, Z_{12242}),         (Z_{11343},Z_{12321},Z_{12144},Z_{12332}, Z_{12215}),\nonumber \\
     &&\hskip .5cm   (Z_{12125},Z_{12142},Z_{12312},Z_{12313}, Z_{12323}), 
        (Z_{11345},Z_{12341},Z_{12344},Z_{12335}, Z_{12245}),\nonumber \\
  &&\hskip .5cm        (Z_{12145},Z_{12342},Z_{12315},Z_{12343}, Z_{12325}),
        Z_{12345} \biggl).
\eea
  Then after  some
 algebras we deduce from (\ref{xxx})  the explicit transformation
\bea
&&Z_{11111}=
\{1+q_1(1,1,1,1,1)+q_1(1,1,1,1,1)+q_1q_2(1,1,1,1,1)+q_1q_2(1,1,1,1,1)\nonumber \\
&&\hskip .7cm+q^2_1(1,1,1,1,1)+q^2_1(1,1,1,1,1)
 +q_1s(1,1,1,1,1)+q^2_1q_2(1,1,1,1,1)+q_1q_2t\},\nonumber  \\
&&Z_{11115}=
\{1+(-1,q_1,q_1,q_1,-1)+(-1,q_1,q_1,-1,q_1)-q_2(1,-q_1,-q_1,1,1)\nonumber \\
&&\hskip .7cm -q_2(1,-q_1,1,-q_1,1)
 -q_1(1,-q_1,1,1,1)-q_1(1,-q_1,1,1,1) +s(1,-q_1,1,1,1)\nonumber \\
&& \hskip .7cm -q_1q_2(1,1,1,1,1)-q_2t\}, \nonumber \\
&&Z_{11144}=
\{1+(q_1,-1,q_1,q_1,-1)-(1,1,-q_1,1,1)-q_2(1,1,-q_1,1,1)-q_2(1,1,1,1,1)\nonumber \\
&&\hskip .7cm +(-q^2_1,-q_1,-q_1,q^2_1,1)+(1,-q_1,-q_1,1,q^2_1)
-(p,s,s,p,p)\nonumber \\
&& \hskip .7cm +q_2(-q_1,1,q_1q_2,1,-q_1)-q_2u\},\nonumber \\
&&Z_{11145}=
\{1-(1,1,-q_1,-q_1,1)-(1,1,-q_1,1,1)+(2,-q_2,q_1q_2,-q_2,2)-(-2,q_2,q_2,q_2,q_2)\nonumber \\
&&\hskip .7cm-(q_1,q_1,q_1,q_1,-1)-(-1,q_1,q_1,-1,q_1)
+(h,-s,-s,h,h) \nonumber \\
&&\hskip .7cm +(2q_1,q_2,-q_1q_2,q_2,2q_1)+2(t+q)\},\nonumber \\
&&Z_{11335}=
\{1-(1,-q_1,1,-q_1,1)-(1,1,1,1,1)-q_2(1,1,1,1,1)+(2,-q_2,2,-q_2,2)
 \nonumber \\
&&\hskip .7cm +(1,q^2_1,1,-q_1,-q_1)+(-q_1,1,-q_1,1,1)
+(h,-p,h,h,h)-(q_1q_2,q_2,q_2,q_1q_2,2)+2u\},\nonumber \\
&&Z_{11315}=
\{1-(1,1,1,-q_1,1)-(1,-q_1,1,1,-q_1)-(-2,q_2,q_2,q_2,q_2)+(-q_2,-q_2,2,q_1q_2,2)\nonumber \\
&&\hskip .7cm+(1,-q_1,1,-q_1,1)+(1,-q_1,1,-q_1,1)+(e,-s,h,-s,e)+(2q_1,q_2,2q_1,2q_1,q_2)+v\},\nonumber \\
&&Z_{11343}=
\{1-(1,1,1,-q_1,1)-(-q_1,1,1,1,1)-(q_2,-2,q_2,q_2,-2)-(-2,-2,q_2,q_2,q_2)\nonumber \\
&&\hskip .7cm+(1,-q_1,1,-q_1,1)+(1,1,1,1,1)
+(e,h,-p,h,e)-(q_1q_2,2,-q_2,-q_2,2)+v+3q\},\nonumber \\
&&Z_{11345}=
\{1-(1,1,1,-q_1,1)-(1,1,1,1,1)+(2,2,-q_2,-q_2,2)+(2,2,2,-q_2,2)\nonumber \\
&&\hskip .7cm+(1,-q_1,1,-q_1,1)+(1,1,-q_1,1,1)
+(-5,h,h,h,-5)+(2q_1,-2,q_2,q_2,-2)+14-5q\},\nonumber \\
&&Z_{12145}=
\{1-(1,1,1,1,1)-(1,1,-q_1,1,1)+(2,2,-q_2,2,2)+(2,-q_2,2,2,-q_2)\nonumber \\
&&\hskip .7cm+(1,1,1,1,1)+(1,-q_1,1,1,-q_1)
-(q+5,5,5,5,-h)-(2,2,-q_2,2,2)+14-2q\},\nonumber \\
&&Z_{12345}=
\{1-(1,1,1,1,1)-(1,1,1,1,1)+(2,2,2,2,2)+(2,2,2,2,2)\nonumber \\
&&\hskip .7cm+(1,1,1,1,1)+(1,1,1,1,1)
-(5,5,5,5,5) -(2,2,2,2,2)+14\}, \label{t5}
\eea
where $q_m= q-m, \>e=q-5,\> h=2q-5, \> s=q^2-5q+5, \> p=q^2-4q+5,\> t=q^2-7q+7,
 \> u=q^2-5q+7, \> v=q^2-10q+14$.
Here, one reads off elements of ${\bf T}_5(q)$ from (\ref{t5}) directly.
For example, the second row of ${\bf T}_5(q)$
has  elements $\{1, (-1, q_1, q_1, q_1, -1), (-1,
\cdots), \cdots,
 -q_2t \}$.  Elements not shown are given by cyclic permutations of
the partition indices within each set of parentheses.
For example, the third row of ${\bf T}_5(q)$ is
$\{1, (-1, -1, q_1, q_1, q_1), (q_1,
\cdots), \cdots,  -q_2t \}$, where  elements
within each set of parentheses are obtained
by cyclically permuting  those in the second row.

\section{The multi-component Potts and chiral Potts models}
The chiral Potts model can be generalized to  more-than-one components.
An $m$-component chiral Potts model \cite{perk,perk1} is a
\be
Q=N_1N_2\cdots N_m
\ee
 state spin model, where $N_\ell, \ell=1,2,\cdots,m$, are
positive integers.   The interaction matrix {\bf U} 
is a $Q\times Q$ matrix  in the form of a direct
product of $m$ cyclic matrices, with elements indexed by
$U(\x_1,\cdots,\x_m\> \vert \> \x_1',\cdots,\x_m')$, $\x_\ell,
\x_\ell' = 1,2,\cdots, N_\ell$,
$\ell=1,\cdots, m$, satisfying  the cyclic property
\be
U(\x_1,\cdots,\x_m \>\vert\>\x_1',\cdots,\x_m') =
  u(\x_1 - \x'_1,\> \cdots,\>\x_m - \x'_m).  \label{mchiral}
\ee
The $m=N_1=N_2=2$ version is  known as 
the Ashkin-Teller model  \cite{at}.
  
It is convenient to introduce
vectors ${\bf \x}=(\x_1,\cdots,\x_m )$ and ${\bf \e}=(\e_1, \cdots,\e_m)$ 
and their scalar product
\be
{\bf \x}\cdot{\bf \e}=\x_1 \e_1/N_1+\cdots+\x_m \e_m/N_m,
\quad \quad \x_\ell, \e_\ell=1,\cdots,N_\ell.
\ee
Then the dual model has the Boltzmann factor
 \be
\l({\bf \e} )={1\over \sqrt Q} \sum^{N_1}_{\x_1=1} \cdots \sum_{\x_m=1}^{N_m}
e ^{2\pi i {\bf \x}\cdot{\bf \e}} \>
u({\bf \x}),
\ee
and the  duality relation (\ref{duality}) now reads
\be 
Q^{-N/2}Z[u({\bf \x})] = Q^{-N^*/2} Z^{(D)}[u^*(\bf \e)].
 \label{mduality}
\ee
Here, again, $\sqrt Q u^*(\bf \e)$ are the eigenvalues of the interaction
matrix {\bf U}.

 Proceeding in a similar fashion as before, we arrive at  
the following duality relation in place of (\ref{gen01}),
\be
Z_{{\bf \x}_1 {\bf \x}_2\cdots{\bf \x}_n}= C_n(Q) 
\!\!\!\sum_{{\bf \e}_{1}{\bf \e}_{2}\cdots {\bf \e}_{n}} 
{ M}({\bf \x}_1, {\bf \x}_2,\cdots,{\bf \x}_n \>
| \> {\bf \e}_{1}, {\bf \e}_{2},\cdots,{\bf \e}_{n})
Z^*_{{\bf \e}_{1} {\bf \e}_{2}\cdots{\bf \e}_{n}}. \label{gen02}
\ee
Here, ${\bf M}$ is a $Q^n\times Q^n$ matrix with elements
 \be
{ M}({\bf \x}_1, {\bf \x}_2,\cdots,{\bf \x}_n \>
|\> {\bf \e}_{1}, {\bf \e}_{2},\cdots,{\bf \e}_{n})
=
\exp \biggl[ 2\pi i ({\bf \x}_{n1}\cdot{\bf \e}_{1}+
{\bf \x}_{12}\cdot{\bf \e}_{2}+\cdots+{\bf \x}_{(n-1)n}\cdot{\bf \e}_{n})\biggr],
\label{melement}
\ee
satisfying  the reciprocal inversion relation
 \be
{\bf M}^2({\bf \x}_1, {\bf \x}_2,\cdots,{\bf \x}_n \>
|\> {\bf \x}'_{1}, {\bf \x}'_{2},\cdots,{\bf \x}'_{n})
=Q^{n+1} \d({{\bf \x}_1,{\bf \x}'_{2}})\d({{\bf \x}_2,{\bf \x}'_{3}})
\cdots\d({{\bf \x}_n,{\bf \x}'_{1}}).
\ee
Defining the $n$-point correlation functions
\bea
P_n({\bf \x}_1,{\bf \x}_2,\cdots,{\bf \x}_n)&=&
 Z_{{\bf \x}_1{\bf \x}_2\cdots {\bf \x}_n}/Z(u),\nonumber \\ 
p_{{\bf y}_1{\bf y}_2\cdots{\bf y}_n}&=&  Q
Z^*_{{\bf \e}_{1} {\bf \e}_{2}\cdots{\bf \e}_{n}}/Z^{(D)}(u^*),
\eea 
then (\ref{gen02}) becomes
\be
P_n({\bf \x}_1,{\bf \x}_2,\cdots,{\bf \x}_n)= { Q^{-(n+1)}} 
\!\!\!\sum_{{\bf \e}_{1} {\bf \e}_{2}\cdots {\bf \e}_{n}} 
{ M}({\bf \x}_1, {\bf \x}_2,\cdots,{\bf \x}_n \>
| \> {\bf \e}_{1}, {\bf \e}_{2},\cdots,{\bf \e}_{n})
p_{{\bf \e}_{1} {\bf \e}_{2}\cdots{\bf \e}_{n}}. \label{gen04}
\ee
This is the most general 
correlation duality relation for the $m$-component chiral Potts model.
An immediate consequence of (\ref{gen04}) is, in analogous to
(\ref{g1}) and (\ref{gamman}), the following expression for the
$n$-point correlation
\be
\Gamma_n \equiv Q^nP_n({\bf \x}, {\bf \x},\cdots,{\bf \x}) -1
= Q^{-1} \>\> 
\!\!\!\sum_{{\bf \e}_{1} {\bf \e}_{2}\cdots {\bf \e}_{n}} 
p_{{\bf \e}_{1} {\bf \e}_{2}\cdots{\bf \e}_{n}} -1.
\ee

{\it The $m$-component Potts model}:
The $m$-component chiral Potts model reduces to   an $m$-component 
(standard)  Potts model
when the interaction (\ref{mchiral}) assumes the form (\ref{standardpotts}) for each
component, namely,
\be
   u(\x_1,\> \cdots,\>\x_m ) =
\prod_{\ell =1}^m \biggl[1+(e^{K_\ell} -1) \d_{N_\ell}(x_\ell, 0)\biggr], \quad
x_\ell = 1, 2, \cdots, N_\ell  ,
\ee
where
\bea
\d_{N_\ell}(\x, \x') &=& 1, \hskip 1.5cm \x=\x' \hskip0.5cm ({\rm mod}\>\>N_\ell), \nonumber\\
  &=& 0, \hskip 1.5cm \x\not=\x' \hskip0.5cm ({\rm mod}\>\>N_\ell). 
\eea
The $m=2$ version is known as the $(N_1, N_2)$ model \cite{domany}.

 Now each ${\bf x}_i$ in (\ref{gen04}) 
is an $m$-component vector.  Let its  components be
$x_{i\ell}=1,\cdots,N_\ell,\> i=1,\cdots,n; \ell=1,\cdots,m$, and let  
$X_\ell$ denote the partition of the 
$n$ integers $1,2,\cdots,n$   dictated by $\{x_{i\ell},\>
i=1,\cdots,n\}$.  
Then,
in analogous to 
 (\ref{relation}), we 
write the partially ordered mappings
 \bea
P_n({\bf \x}_1,{\bf \x}_2,\cdots,{\bf \x}_n)
&=&\sum_{X_\ell}
A_{X_1X_2\cdots X_m}\d_{N_1}(X_1)\d_{N_2}(X_2)\cdots\d_{N_m}(X_m), \nonumber \\
p_{{\bf y}_1{\bf y}_2\cdots{\bf y}_n}
&=&\sum_{Y_\ell }
B_{Y_1Y_2\cdots Y_m}\d_{N_1}(Y_1)\d_{N_2}(Y_2)\cdots\d_{N_m}(Y_m).\label{mrelation}
\eea
  With these notations, we have the following corollary
of our Theorem:

\noindent
{\it Corollary}:

For the $m$-component Potts model whose correlation functions are
 (\ref{mrelation}), we have
\bea
A_{X_1X_2\cdots X_m}&=&\prod^m_{\ell=1}N_\ell^{-|X_\ell|}B_{Y_1Y_2\cdots Y_m},
\hskip 1.2cm {\rm for \>\>planar\>\>}X_\ell \leftrightarrow Y_\ell,
   \nonumber  \\
   &=&0,\hskip 4.2cm{\rm otherwise}. \label{coroll}
\eea
Here, explicitly, by using
 \bea
Z_{X_1X_2\cdots X_m}/Z&=&\sum_{X'_\ell\preceq
X_\ell}A_{X'_1X'_2\cdots X'_m}\nonumber \\
B_{Y_1Y_2\cdots Y_m}&=&\sum_{Y_\ell'\preceq Y_\ell}
p_{Y'_1Y'_2\cdots Y'_m}\prod^m_{\ell=1}
\mu(Y_\ell',Y_\ell),
\eea
in analogous to (\ref{xxx})
  one can  reduce the  correlation duality relation (\ref{gen02})
to
 \bea
Z_{X_1X_2\cdots X_m}&=&C_n(Q)\sum_{X'_\ell\preceq
X_\ell}\sum_{Y_\ell''\preceq Y_\ell'}Z^*_{Y''_1Y''_2\cdots Y''_m}  \nonumber \\
&& \hskip 1cm \times 
 \prod^m_{\ell=1}
N_\ell^{|Y_\ell'|}\mu(Y_\ell'',Y_\ell'),\hskip 1cm
X_\ell' \leftrightarrow Y_\ell'.  \label{mpottsdual}
\eea
This is the desired duality relation 
for the $m$-component Potts model. 
Now there are  $(c_n)^m$ partial
partition functions $Z$ and $Z^*$.  Considering $Z$ and $Z^*$ as  tensors 
 ${\bf Z}_{m}$ and ${\bf Z}^*_{m}$ of rank $m$, one can rewrite (\ref{mpottsdual}) 
more compactly as
\begin{equation}
{\bf Z}_{m}=QC_n(Q)\biggl[{\bf T}_n(N_1)\otimes \cdots \otimes{\bf T}_n(N_m)  
 \biggr] \cdot{\bf Z}^*_m,  \label{mmm}
\end{equation}
where the tensor products are over $m$
$ c_n\times c_n$ matrices ${\bf T}_n(q)$  defined in (\ref{tn}).
The last expression  generalizes the $m=2$ results for the $(N_1, N_2)$ model 
reported in \cite{wulu}.

\section{Summary and acknowledgement}

We have considered the chiral Potts model and obtained the duality relation for
its correlation functions of $n$ sites on the boundary
of a planar lattice.  The result is
given by (\ref{gen01}) and (\ref{gen03}).  By specializing this result to
the standard Potts model, we establish a recent conjecture of \cite{wuhuang}
on the correlation duality which we now state in (\ref{conjecture})
as a Theorem.  
  The explicit duality relation  relating the
partial partition functions is  given in (\ref{xxx}), with 
the $n=2,3,5$ expressions explicitly worked out.
The formulation is next extended to the multi-component chiral
Potts model, leading to    the correlation duality relations
(\ref{gen02}) and (\ref{gen04}). 
Again, specializing the results to the $m$-component Potts model,
we obtain the Corollary (\ref{coroll}) and the
correlation duality relations 
(\ref{mpottsdual}) and (\ref{mmm}). 

 Work has been supported in part by NSF Grant DMR-9614170.

\newpage
\begin{center}

{\bf Figure captions}

\end{center}

\noindent
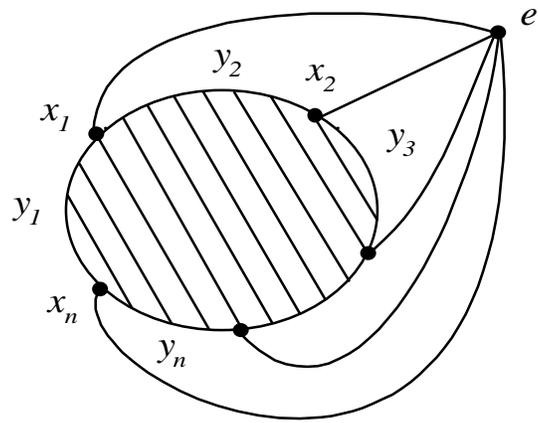
Fig. 1.
 A planar graph \L and $n$ sites $i,j,\dots m,\ell$ on the boundary.

\medskip
\noindent
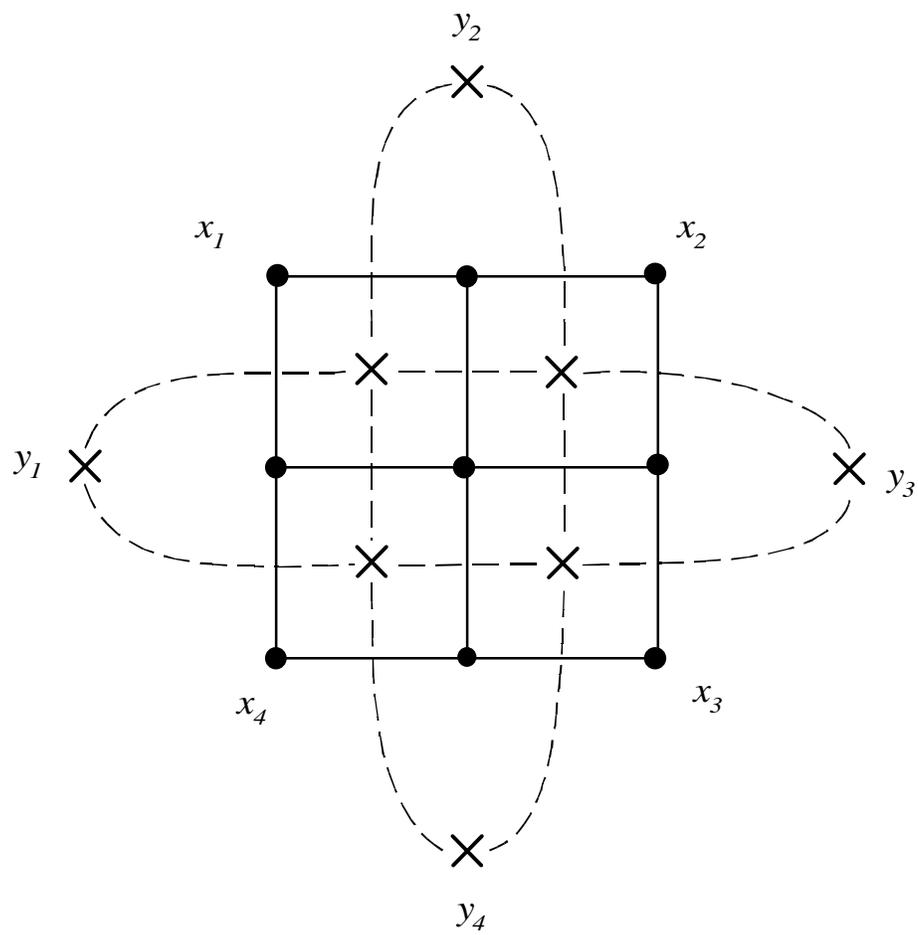
Fig. 2. Reciprocal graphs \L (solid lines) and ${\cal L}^*$ (broken lines).
 
\medskip
\noindent
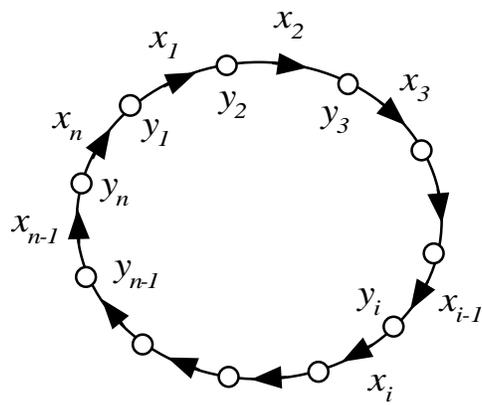
Fig. 3. Graphical representation $G_n$ of the matrix element (\ref{melement}). 

\medskip
\noindent
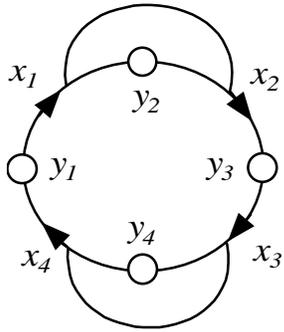
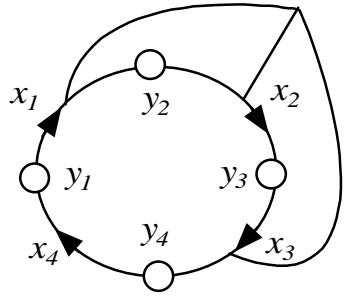
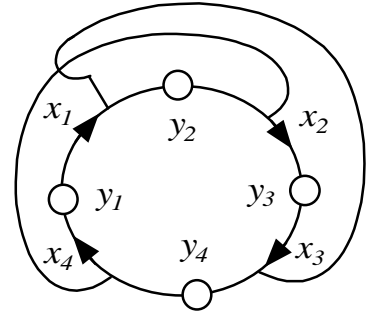
Fig. 4.  Examples of connectivities on $G_4$. (a) $X= \{1133\}$. (b) $X= \{1114\}$.
(c) $X= \{1212\}$.

\medskip
\noindent
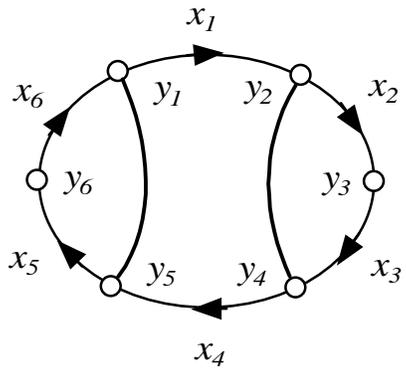
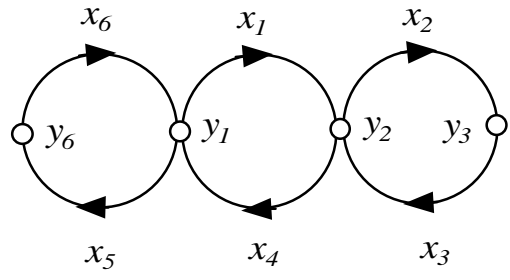
Fig. 5. The contraction of the the $ Y=\{123216\}$ graph
of $G_6$  into a cactus tree.

\end{document}